\documentclass[twocolumn]{aastex63} 
\usepackage{amsmath}
 
\shorttitle{} 
\shortauthors{} 
 
\begin{document} 
 
\received{} 
\accepted{} 
 
\title{Tidal Asteroseismology: Possible Evidence of Non-linear Mode Coupling in an Equilibrium State in Kepler Eclipsing Binary KIC 3230227}  
 
\author{Zhao Guo} 
\affil{Center for Exoplanets and Habitable Worlds, 525 Davey Laboratory, The Pennsylvania State University, University Park, PA 16802, USA} 
\affil{Department of Astronomy \& Astrophysics, 525 Davey Laboratory, The Pennsylvania State University, University Park, PA 16802, USA}
 

 
\begin{abstract} 
Previously, a series of tidally-excited oscillations were discovered in the eccentric eclipsing binary KIC 3230227. The pulsation amplitudes and phases suggest the observed oscillations are prograde quadruple modes. In this paper, we refine the analysis and extract more oscillation frequencies. We also study the temporal variations of amplitudes and phases and show that almost all modes have stable phases and amplitudes. We then focus on the non-orbital-harmonic oscillations. We consider two formation mechanisms: 1) nonlinear response of the surface convective layer, and 2) nonlinear three/multi-mode coupling. Although the former can explain some of the observed features, we find the latter mechanism is more probable. Assuming that these are coupled modes, the constant amplitude/phase over four years can be explained by either an equilibrium state in the mode coupling or modes undergoing limit cycles with very long periods.  The observed frequency detuning and the calculated damping rates of the daughter modes favor the equilibrium-state interpretation. This is verified by integrating the amplitude equations of three-mode coupling. We find that the steady-state relation derived in Weinberg et al., which relates the observed frequency detuning, phase detuning, and mode damping rates, is approximately satisfied for one mode triplet. We also try to identify the azimuthal number of the modes based on the observed mode amplitude ratios and the selection rules in nonlinear three-mode coupling. We discuss further implications of these observations on nonlinear tidal asteroseismology.

\end{abstract} 
 
 
 

\section{Introduction}                               

In addition to the free oscillations excited by internal sources (opacity, turbulence, etc.), stellar oscillation can also be driven by the tidal force from a companion.
Tidally Excited Oscillations (TEOs) have been observed in many eccentric binary systems (Welsh et al. 2011; Fuller \& Lai 2012; Kirk et al.\ 2016; Hambleton et al.\ 2016, 2018; Guo et al.\ 2017, 2020). They are direct manifestations of dynamical tides. In the linear theory, stars are expected to ring at the frequencies of the driving force, which are multiple integer times of the orbital frequency.  This is indeed in agreement with observations. However, in the nonlinear regime, stars can ring at non-orbital-harmonic frequencies, and this has been observed in KOI-54. Burkart et al.\ (2012) and O'Leary \& Burkart (2014) studied the anharmonic modes of this star in the framework of three- and multiple-mode coupling.

Three-mode coupling is the lowest order nonlinear effect, and it arises naturally in the equation of motion if the second-order perturbation terms are kept. This has been discussed in detail in the literature (Dziembowski 1982; Schenk et al.\ 2002; Wu \& Goldreich 2001; Pnigouras \& Kokkotas 2015).

Recent space observations of stellar oscillations show increasing evidence of nonlinear effects. There have been many studies on the nonlinear mode coupling, including, but not limited to:

1. The three-mode coupling, as a leading amplitude limitation mechanism, has been studied for $\delta$ Scuti stars (Dziembowski  \& Krolikowska 1985; Dziembowski et al.\ 1988). It is found that parametric resonance can limit the pulsation amplitude to the observed level and also explain the fact that only one-third of stars in the instability strip show detectable pulsations. Lee (2012) studied this mechanism for low-frequency g-modes in slowly pulsating B-stars and found the fractional amplitudes of the radiative luminosity are confined to be $10^{-4}-10^{-3}$ and r-modes play an important role in the coupled mode triad. Wu \& Goldreich (2001) showed that the parametric instability threshold agrees with the observed pulsation amplitude of DAV white dwarfs. For solar-like oscillating stars, Kumar \& Goldreich(1989), Kumar et al.\ 1994, and  Lavely (1990) show that three-mode coupling cannot effectively damp the over-stable p-modes. The nonlinear mode coupling can significantly increase the tidal dissipation in solar-type exoplanet host stars (Weinberg et al.\ 2012; Essick \& Weinberg 2016). For evolved oscillating red giants,  Weinberg \& Arras (2019) examined the nonlinear damping of mixed dipole modes and showed that at least for upper red giants, strong nonlinearity plays a significant role in the mode damping in the core.

2. Amplitude, frequency, and phase variations have been studied in several compactor pulsators (sub-dwarf B-stars and white dwarfs, Zong et al.\ 2016) and hundreds of $\delta$ Scuti stars (Bowman et al.\ 2016). Besides other mechanisms such as beating and binarity, nonlinear mode coupling is a primary reason for these variations. In fact, amplitude equations (AEs) have been used to account for various kinds of resonant mode couplings (Van Hoolst \& Smeyers 1993; Goupil \& Buchler 1994; Van Hoolst 1994; Buchler et al.\ 1997).

3. Kurtz et al.\ (2015) show that in g-mode pulsators ($\gamma$ Doradus, Slowly Pulsating B-stars, and Be stars), combination frequencies can dominate the Fourier spectrum and their amplitudes can be higher than the principle frequencies. Combination frequencies are also common in p-mode pulsators such as $\delta$ Scuti stars (Breger \& Montgomery 2014; Balona 2016). They can be explained by nonlinear mode coupling (see also Saio et al.\ 2018). Note that the nonlinear response of the stellar atmosphere can also generate combination frequencies (Wu 2001). This mechanism would be elaborated in Section 3.3.

4. Some DAV-type pulsating white dwarfs show sporadic outbursts (Bell et al.\ 2015; Hermes et al.\ 2015). This has been explained successfully by limit cycles of daughter modes undergoing three-mode coupling (Luan \& Goldreich 2018).

Guo et al.\ (2017, G17 hereafter) performed a binary modeling the Kepler eclipsing binary KIC 3230227 and derived the following fundamental parameters for the two components: 
$M_1 = 1.84\pm 0.18M_{\odot}$, $M_2 = 1.73\pm0.17M_{\odot}$ $R_1 = 2.01 \pm 0.09 R_{\odot}$, $R_2 = 1.68 \pm 0.08R_{\odot}$, as well as orbital parameters: $P_{\rm orb}=7.047106\pm 0.000018$ days, eccentricity $e=0.600 \pm 0.05$, argument of periastron $w_p=293^{\circ}\pm 1^{\circ}$, and orbital inclination $i=73.42^{\circ}\pm 0.27^{\circ}$. They also compared the theoretical amplitudes and phases of the ten dominant TEOs and found they agree with $l=2, m=2$ prograde modes.

In Section 2, We adopt a more strict noise model in the Fourier spectrum and report additional significant frequencies that are likely tidally excited. 
In Section 3, we first examine the amplitude and phase variations of these modes. Then we focus on the mechanism that can explain the three non-orbital-harmonic modes: nonlinear mode coupling and nonlinear response of the convective layer. In the last section, we discuss the limitation and caveats of this work and future prospects.

\section{Orbital harmonic TEOs} 

We performed the standard pre-whitening procedure (Lenz \& Breger 2005) to extract significant oscillation frequencies.
The extracted frequencies with $S/N \ge 10$ are shown in the upper panel of Figure 1 and listed in Table 1. We model the noise in the Fourier spectrum with a Lorentzian-like function. This function has been used to model intermediate and massive stars (Pablo et al.\ 2017; Bowman et al.\ 2019; Handler et al.\ 2019). It is a more strict noise model compared to the empirical smoothing method in G17. The 10 dominant frequencies reported in G17 are marked with symbols in color. We find the additional frequencies also have constant amplitudes and phases similar to the 10 previously reported frequencies.

The lower panel of Figure 1 shows pulsation phases of all frequencies. Some of the newly extracted frequencies are likely $|m|=2$ prograde modes, and they have orbital-harmonic numbers $N=f/f_{orb}$ between 22 and 30 ($f=f_{11}, f_{13}, f_{15}, f_{20}, f_{18}, f_{25}$), similar to the previously reported $f_{5},f_{6},f_{7},f_{8},f_{9},f_{10}$ in G17, so is $f_{22}$ (with N=5). Note that whether modes are prograde or retrograde cannot be determined from the phases alone. The labeling ($m=2$ or $m=-2$) in Figure 2 is only for the convenience of calculating the actual theoretical phases of TEOs and they must be interpreted as $|m|=2$.

\section{Formation Mechanism for the Non-orbital-harmonic TEOs}  

\subsection{Have the modes settled into an equilibrium state?}
We focus on the three modes that are not multiple integer times of orbital frequency: $f_1=9.88f_{orb}, f_2=12.12f_{orb}$, and $f_4=13.88f_{orb}$.
G17 already alluded that the anharmonic TEOs are likely due to nonlinear resonant mode coupling. These modes  satisfy the relation that the sum of two daughter-mode frequencies is equal to the frequency of the parent mode: ($f_b+f_c \approx f_a$). The observed two triplets indeed follow this relation: (Triplet 1: $f_{2}+f_{1} \approx f_{11}\approx 22.0f_{orb}$) and (Triplet 2: $f_{  4}+f_{2} \approx f_{15}\approx 26.0f_{orb}$). The two triplets share one daughter mode $f_2$, but here we assume these frequencies can be treated as two independent triplets and adopt the theory of three-mode coupling to explain the observations.
The caveat of this assumption is discussed in Section 4.

Figure 2 illustrates the variations of pulsation amplitudes (upper) and phases (lower). They are calculated by performing Fourier analysis of light curves in a running window with a width of 50 days. The daughter modes are represented by open circles and the parent modes in filled squares.
Note that in the lower panel, the two parent modes have similar phases ($\phi_{11}, \phi_{15} \approx 0.33$).
We also show the sum of the two daughter-mode phases in thick dashed lines ($\phi_4 + \phi_2$ and $\phi_1 + \phi_2$). It 
can be seen that, compared to their parents' phases $\phi_{15}$ and $\phi_{11}$, they satisfy the following relations:  $\Delta\phi=\phi_{15}-(\phi_4+\phi_2) \approx -0.2$  and $\Delta\phi=\phi_{11}-(\phi_2+\phi_1) \approx 0 $. 
Both amplitudes and phases are essentially stable over the {\it Kepler} observation of four years.  This motivates us to examine if the coupled modes have reached a non-linear equilibrium state.

Generally, a three-mode coupling system can be described by the amplitude equations. The AEs can be derived by keeping the $2^{nd}$-order perturbation terms in the equation of motion or by a Hamiltonian formalism (Kumar \& Goldreich 1989; Wu \& Goldreich 2001; Schenk et al. 2002). In the context of parametric instability, the parent mode (a) is excited and becomes unstable. Its amplitude exponentially increases and becomes nonlinear. When its amplitude surpasses a threshold, the parametric instability threshold, the parent mode can transfer energy to two daughter modes (b and c). The amplitude equations of the three-mode system read:

\begin{equation}
\begin{split}
\dot{q_a}+(i\omega_a + \gamma_a)q_a=i\omega_a 2 \kappa q_b^* q_c^*\\
\dot{q_{b}}+(i\omega_{b} + \gamma_{b})q_{b}=i\omega_{b} 2 \kappa q_{a}^* q_{c}^* \\
\dot{q_{c}}+(i\omega_{c} + \gamma_{c})q_{c}=i\omega_{c} 2 \kappa q_{a}^* q_{b}^*,
\end{split}
\end{equation}
where $q_{i} (t)$ are complex mode amplitudes, $\gamma_i$ are linear damping rates, and $\kappa$ is the mode coupling coefficient which depends on the mode eigenfunctions and their normalization.  When explicitly written as real amplitudes $A_{i}$ and real phases $\delta_{i}$: $q_i=A_ie^{i\delta_i}\  (i=a,b,c)$,
the AEs in the end  become differential equations of $A_{a}, A_{b}, A_{c}$, and $\delta$ with parameters $\gamma_a, \gamma_b, \gamma_c, \kappa$. The phase information is only related to the phase detuning parameter $\delta=\delta_a+\delta_b+\delta_c$. The signs of $\omega_i$ ($i=a, b, c$) can be chosen so that the three modes satisfy the approximate resonance condition ($\delta \omega=\omega_a+\omega_b+\omega_c \approx 0$). Depending on the parameters $\gamma_i$ and frequency detuning $\delta \omega$, the system can behave in different ways:  steady states, limit cycles, chaos, or unstable exponentially growth (Wersinger et al.\ 1980; Moskalik 1985).

In the context of KIC 3230227, the parent mode is likely driven by a linear dynamical tide (an orbital harmonic, $U_a(t)=U_ae^{-i\omega t}$) and not self-excited. The AE of the parent mode has an extra term on the right side: $ i\omega_a U_a e^{-i\omega t}$. Removing the time derivative of eq. 1,  W12  derived the non-linear equilibrium solution. The mode amplitudes of two daughters in the equilibrium are related by their quality factors ($\omega_i/\gamma_i$):
\begin{equation}
\frac{A_b^2}{A_c^2}=\frac{\omega_b/\gamma_b}{\omega_c/\gamma_c} \\
\end{equation}

In addition, the phase detuning $\delta$ and the parameter $\mu$ (relating the frequency detuning and mode damping rates of the daughter modes) satisfy the following algebraic relation: 

\begin{equation}
\begin{split}
\tan(2\delta)=-2\mu/(1-\mu^2) \\
LHS \ \ \ \ \ \ \ \ \ \ \ \ \ \ \ \ \ \ \  RHS
\end{split}
\end{equation}
where $\mu=(\Delta_b+\Delta_c)/(\gamma_b+\gamma_c) $. $\Delta_b+\Delta_c=\omega-\omega_b-\omega_c$ is the frequency detuning parameter. We adopt the convention that $\gamma_b, \gamma_c >0$ (damped), $\gamma_a <0$ (excited), and the signs of $\omega_b, \omega_c, \delta_b, \delta_c$ are negative (Lee 2012). Note that since the observed light curve is modeled as a sum of sinusoidal functions $\sum_i a_i\sin(2\pi(f_it+\phi_i))$, there is a phase offset for the phase detuning parameter $\delta=\pi/2 -\phi$.

In the following, we examine closely if equation (3) is satisfied in the two observed triplets. All the observables of the two triplets are summarized in Table 2. Note that although we use angular frequency $\omega_i$ in the above equations (1) and (2), we switch to linear frequency $f_i$ and damping rates $\gamma_i$ ($i=a, b, c$) below when dealing with observables.  Given the context, the meaning is not ambiguous.

To calculated the mode damping rates, we obtained a stellar model with MESA evolution code (Paxton et al.\ 2011, 2013, 2015) which has the observed parameters of the primary star in KIC 3230227 ($M_1=1.84M_{\odot}$, $R_1=2.01R_{\odot}$, Z=0.02). Assuming a stellar rotation rate of two thirds of the pseudo-synchronous rate $f_{rot}=(2/3)\times 4.1f_{orb}=0.3879$ day$^{-1}$ (Hut 1981; Zimmerman et al.\ 2017),  we then calculate the non-adiabatic eigen-frequencies and eigen-functions of $l=2, m=0$, $m=1, 2$ (prograde) g-modes with the GYRE oscillation code (Townsend \& Teitler 2013). The effect of rotation is implemented in the traditional approximation (Unno et al.\ 1989; Bildsten et al.\ 1996; Lee \& Saio 1997; Townsend 1997).
Figure 4 (lower panel) shows the linear damping rates ($\gamma$) of these modes.
The range of $\gamma$ for the three daughter modes ($f_1, f_2, f_4$) lies between about $10^{-7}$ and $10^{-4}$ day$^{-1}$. 

We then use the above mode damping rates and the observables $\delta$ and $\Delta_b+\Delta_c$ to examine the equilibrium-state relation in eq. (3).
For Triplet 1:, we find the phase detuning $\delta=0.234 \pm 0.021$ (in $2\pi$) and the frequency detuning $\Delta_b+\Delta_c=−0.0001687(86)$ (day$^{-1}$); for the second triplet, we find $\delta=0.438 \pm 0.029$ and $\Delta_b+\Delta_c=-0.0001900(97)$. 
The upper panel of Figure 4 shows the difference between the right-hand side (RHS) and the left-hand side (LHS) of eq.\ (3). For the two observed mode triads, the difference is plotted as a function of the sum of the two daughter-mode damping rates ($\gamma_b+\gamma_c$). 
We can see that for a large range of $\gamma_b+\gamma_c$, Triplet 1 almost satisfies eq.\ (3), especially around $\gamma_b+\gamma_c \approx 10^{-5}$ where $|LHS-RHS|$ equals zero. However, it is more difficult for Triplet 2 to satisfy eq.\ (3), except for a narrow window around $\gamma_b+\gamma_c \approx 10^{-4}$. The damping rates $\gamma_b+\gamma_c $ when eq.\ (3) is approximately satisfied for the triplet 1 are about the same order of magnitude ($10^{-5}-10^{-4}$) with the damping rate from GYRE calculations (Figure 4, lower panel). Given that the mode amplitudes and phases are almost constant over the four-year observation, it is possible that the modes in one or both triplets have settled to a steady state or are undergoing long-term variations.

According to Wu \& Goldreich (2001), the resonance three-mode coupling (parametric instability) leads to a steady state if $|\delta f|> \gamma$, and to limit cycles if  $|\delta f|< \gamma$ ($\gamma$ is the daughter-mode damping rate, $\delta f$ is the frequency detuning of the triplet). For Triplet 1, the frequency detuning $\delta f=f_1+f_2-f_{11}=0.00016$ day$^{-1}$. The daughter-mode damping rates ($\gamma_b, \gamma_c$) depend on the mode identification. From Figure 4, the most probably range of ($\gamma_b, \gamma_c$) is between $10^{-6}$ and $10^{-5}$ day$^{-1}$ (see Sec. 3.2). These typical damping rates are generally smaller than $\delta f$, and the steady-state condition is satisfied. For Triplet 2, the corresponding frequency detuning is $\delta f=f_2+f_4-f_{15}=0.00023$ day$^{-1}$, and again it is generally larger than the daughter-mode damping rates. In fact, the equilibrium state is a stable attractor for mode triplets with $|\delta f| > \gamma$.  Without scanning the parameter space of amplitude equations, we cannot completely rule out the possibility that the modes are undergoing a limit cycle with a period much longer than the observed time span (4 yr). The timescale of limit cycles for the g modes in question can indeed be quite long. But the frequency detuning and the daughter-mode damping rates seem to favor the equilibrium state interpretation.

We experiment to use equation (1), with and without the tidal term $U_a(t)$ in the first line, to study the behavior of coupled modes. We calculate the mode coupling coefficient $\kappa$ by using the expression A55 in Weinberg et al.\ (2012). The modes in question from our MESA stellar structure model yield $\kappa$ values on the order of $1-10$. The AEs are integrated by using the 4th-order Runge-Kutta method. First, without the tidal term, we assume the parent mode is unstable ($\gamma_a <0$). The left panel of Figure 3 shows that, for the g modes listed in Table 2,  the three-mode system ($f_1,f_2,f_{11}$) with the observed frequency detuning $\delta f$ and the calculated mode damping rates $\gamma_b, \gamma_c$ can indeed evolve into an equilibrium state. On the other hand, if the parent mode is stable ($\gamma_a >0$) but instead driven by an orbital harmonic term $U_a(t)$, a representative example for the mode triplet is shown in the right panel of Figure 3. Again, the example shows that a three-mode system such as the g-mode triples in KIC 3230227 can evolve into a steady state. A detailed study of the mode behavior would require exploring the full parameter space of AEs and is beyond the scope of this paper.

\subsection{Mode Identification from the Three-mode-coupling Selection Rules?}
Three-mode coupling needs to satisfy the energy and angular momentum conservation. Thus a series of selection rules must be applied (Schenk et al.\ 2002; Burkart et al.\ 2012; O'Leary \& Burkart 2014):
\begin{equation}
(l_a+l_b+l_c) \ mod \ 2 =0
\end{equation}

\begin{equation}
|l_b-l_c| \le l_a \le l_b+l_c
\end{equation}

\begin{equation}
m_a=m_b+m_c.
\end{equation}

We attempt to use these rules to identify the azimuthal number $m$.
Figure 1 (lower panel) shows that the two parent modes ($f_{11}=22f_{orb},f_{15}=26f_{orb}$) have phases close to the theoretical $l=2, m=2$ modes. 
The modes in question have relatively large amplitudes and it is reasonable to assume that they are all $l=2$ modes.  $l=3$ or higher modes suffer from more significant geometric cancellation as well as a factor of $(R/a)^{l-2}=(0.076)^{l-2}$ decrease in amplitude. Thus selection rules in eq.\ (4) and (5) are already satisfied. There is no signature of spin-orbit misalignment, so we can further assume that the tidally-excited parent modes are $m=2, 0, $ or $-2$ since $m=\pm 1$ modes are not excited by tides\footnote{This argument only applies to the parent modes, which are excited by linear tides. For the daughter modes, which are excited by the parametric instability, they can still be $l=2, m=\pm 1$ modes.}. Assuming the two parents modes are indeed prograde $l=2, m_a=2$ modes, eq.\ (6) implies that $m_b$ and $m_c$ of the two daughter modes can be $m=1, 2$ or $m=0$. Since two triplets share one daughter ($f_2$), the non-sharing daughters in the two triplets must have the same $m$, so we have three scenarios here: (1: $f_1,f_4,f_2$ are $m=0,0,2$ modes, respectively);
(2: $f_1,f_4,f_2$ are $m=2,2,0$ modes, respectively); (3: $f_1,f_4,f_2$ are all $m=1$ modes).

From the lower panel of Figure 4, we can estimate the mode damping rates:

\textbf{Scenario 1}:
if  $f_2$ is an $m=2$ prograde mode, then $\gamma_1 \approx 7.9\times 10^{-6}$, $\gamma_4 \approx 6.1 \times 10^{-7}$, $\gamma_2 \approx 3.6 \times 10^{-5}$, $\gamma_1+\gamma_2 \approx 4.4 \times 10^{-5}$,
and $\gamma_4+\gamma_2 \approx 3.6\times 10^{-5}$ (day$^{-1}$). The upper panel of Figure 4 indicates that triplet 1 approximately satisfies eq.\ (3), and triplet 2 does not;

\textbf{Scenario 2}:
if $f_2$ is an $m=0$ mode, then $\gamma_1=2.9\times 10^{-4}$, $\gamma_4 \approx 1.0\times 10^{-5}$, $\gamma_2 \approx 1.6\times 10^{-6}$, $\gamma_1+\gamma_2 \approx 2.9 \times 10^{-4}$, $\gamma_4+\gamma_2 \approx 1.2 \times 10^{-5}$, neither triplets approximately satisfy the eq.\ (3) relation;

\textbf{Scenario 3}: if  $f_2$ is an $m=1$ prograde mode, then $\gamma_1 \approx 4.4\times 10^{-5}$, $\gamma_4 \approx 2.2\times 10^{-6}$, $\gamma_2 \approx 7.7\times 10^{-6}$, $\gamma_1+\gamma_2 \approx 5.2 \times 10^{-5}$, $\gamma_2+\gamma_1 \approx 9.9\times 10^{-6}$. Again, triplet 1 approximately satisfies eq.\ (3), and triplet 2 does not;

Thus scenarios 1 and 3 seem to be favored. We then examine whether the observed amplitudes ratios the three anharmonic modes can be explained by the non-linear equilibrium relation eq.\ (2).
Following Dziembowski (1977) and Burkart et al.\ (2012), the observed luminosity variation of pulsation modes can be expressed as:

\begin{equation}
\frac{\Delta L_{\alpha}}{L}=A_{\alpha} \left  [(2b_l-c_l)\frac{\xi_{r,\alpha}(R)}{R} +b_l \frac{\Delta F_{\alpha}(R)}{F(R)}\right ]Y_{lm}(i_s,\phi),
\end{equation}
where $A_{\alpha}$ is the intrinsic mode amplitude; the term in the square bracket depends on the surface values of mode eigenfunctions: $\xi_r$ is the radial displacement, $\Delta F/F$ is the Lagrangian flux perturbations, and $b_l$ and $c_l$ are limb darkening coefficients; the last term $Y_{lm}(i_s,\phi) \propto \sqrt{\frac{(2l+1)(l-m)!}{4 \pi(l+m)!}} P^m_l(\cos i)$ accounts for the geometric cancellation when the disk-integration is performed ($i_s $ is the inclination angle between the pulsation axis and the line of sight).

Following Weinberg et al.\ (2012), we normalize the mode eigenfunction so that the modes have unit mode energy (a modified mode orthogonality relation for rotating stars): $\omega_{\alpha}^2 \int{\rho \xi^*_{\alpha} \cdot \xi_{\alpha} dx^3}+\omega_{\alpha} \int{\rho\xi^{*}_{\alpha} \cdot (\Omega \times \xi_{\alpha}) dx^3}=GM^2/R$ (Fuller 2017). By using the GYRE eigenfunctions mentioned above, we then calculate the square bracket term in eq.\ (7) for $(l=2,m=1)$, $(l=2, m=0)$, and $(l=2,m=2)$ modes.

In the Fourier spectrum, the observed luminosity variations ($\Delta L/L$) of the three daughters ($f_1, f_4, f_2$) are: $( (\frac{\Delta L}{L})_{f_1}:(\frac{\Delta L}{L})_{f_4}:(\frac{\Delta L}{L})_{f_2})=(0.179: 0.338: 0.192)$$\approx (0.93:1.76:1.00)$.

If we assume that the daughter modes satisfy the steady-state amplitude-ratio relation (eq.\ (2)), the intrinsic mode amplitude is proportional to the quality factor ($f_i/\gamma_i$). We find that, for
Scenario 3: $(A_1:A_4:A_2)_{intrinsic}=(\sqrt{\frac{f_1}{\gamma_1}}:\sqrt{\frac{f_4}{\gamma_4}} : \sqrt{\frac{f_2}{\gamma_2}})\approx (0.38:2.00:1)$, and
for Scenario 1: $(A_1:A_4:A_2)_{intrinsic}\approx (1.91:8.19:1)$. 

For the square bracket term in eq.\ (7), we find that for Scenario 3:  $[]_{f_1} : []_{f_4}: []_{f_2}=2.265: 0.522: 1$, and for Scenario 1: $[]_{f_1} : []_{f_4}: []_{f_2}=0.806: 0.186: 1.00$. As for the geometric term $Y_{lm}(i_s,\phi)$, for Scenario 3 ($(f_1, f_4,f_2)=({l=2,m=1})$), the three daughters suffer from the same geometric cancellation, $Y_{21}(i_s,\phi): Y_{21}(i_s,\phi),Y_{2,1}(i_s,\phi)= 1:1:1$; for Scenario 1 ($(f_1, f_4,f_2)=(l=2,{m=0,0,2}))$, $Y_{20}(i_s,\phi): Y_{20}(i_s,\phi),Y_{22}(i_s,\phi)= 0.67:0.67:1.0$ (with $i_s=73.42^{\circ}$).

Thus we obtain the amplitude ratios of theoretical luminosity variation for the three daughter modes. For Scenario 3: $( (\frac{\Delta L}{L})_{f1}:(\frac{\Delta L}{L})_{f_4}:(\frac{\Delta L}{L})_{f_2})=0.861: 1.043: 1$, and for Scenario 1:   $( (\frac{\Delta L}{L})_{f_1}:(\frac{\Delta L}{L})_{f_4}:(\frac{\Delta L}{L})_{f_2})= 1.031:  1.023: 1.000$.  It seems that both scenarios can explain the observed amplitude ratios of two daughters ($f_1, f_2$) in triplet 1, but not those in triplet 2  ($f_4, f_2$).  The ratios in Scenario 3 is in slightly better agreement with the observed values: $(0.93:1.76:1.00)$ than Scenario 1.

We cannot determine the mode identification with certainty. As will be elaborated in the next section, the calculations here are based on the assumption that the five-mode system can be approximated by two separate three-mode couplings. It seems that the three modes in Triplet 1 approximately satisfy the three-mode-coupling equilibrium equations (eq. 2, 3), but the Triplet 2 does not. The caveats will be discussed in the final section.

\subsection{Can Combination Frequencies Arising from the Nonlinear Response of the Stellar Atmosphere Explain the Observed Mode Triplets?}

As the pulsations passing through the stellar atmosphere, the convective layer can respond nonlinearly and produce combination frequencies (Brickhill 1992; Winget et al.\ 1994). Wu (2001) derived the analytical expression for the combination frequencies and successfully used this mechanism to explain the observations of pulsating white dwarfs. In this mechanism, two sinusoidal signals (e.g., flux perturbations) with frequencies ($f_i, f_j$) incident upon the bottom of the convection zone. Due to the non-linear response of the convective layer, the signals are non-linearly mixed, and the emergent flux perturbations can be comprised of sinusoidal signals with frequencies $f_i$, $f_j$, $f_{i}+f_{j}$, $f_{i}-f_{j}$, $2f_{i}$, $2f_{j}$, etc. For this mechanism to work, the convective turn-over timescale in the surface convection zone should be much shorter than the pulsation period. For an A-type main-sequence star such as the primary in KIC 3230227, we find that the convective turn-over timescale in the thin surface convective layer near the stellar surface can indeed satisfy this condition. Thus it seems to be a viable mechanism to produce the observed combination frequencies in KIC 3230227. 

Although this mechanism can generate combination frequencies for an A-star, we argue that it is not very likely to explain the observed g-mode triplets here. The observed triplets are in the form of $f_b+f_c \approx f_a$, where $f_a$ is an orbital harmonic, and $f_b, f_c$ are non-orbital-harmonic frequencies. For this mechanism to work, it would need two incident non-orbital-harmonic signals ($f_b, f_c$) at the bottom of the surface convective layer, so that the emergent signal can have an orbital harmonic. But how would these two non-orbital-harmonic modes be excited to large amplitudes? Dynamical tide, to the linear order, should only excited orbital harmonic frequencies. It is less likely that two non-orbital-harmonic modes are directly driven by the non-linear tide to large amplitudes. What is more likely is that, an orbital-harmonic mode is driven to a large amplitude by the linear tide and then suffers from parametric instability. This produces two non-orbital-harmonic daughter modes. If the two non-orbital-harmonic signals are instead self-excited g modes such as those in $\gamma$ Dor stars, then it is just a coincidence that their sum is an orbital harmonic. This is possible, but if we examine a similar system KOI-54: there are many daughter pairs whose frequencies satisfy the same condition $f_b+f_c=f_a=91f_{orb}$ (O'Leary \& Burkart 2014). These daughter pairs are too numerous to be purely explained by chance. Instead, three or multi-mode coupling can naturally explain this: these are daughter pairs coupled to the same parent mode.

\section{Discussions and Future Prospects}
In KOI-54, the daughter modes have smaller amplitudes than their parents. However, in KIC 3230227, the two daughters in both triplets have much larger magnitudes. The nonlinear effect is thus a significant factor that cannot be ignored in interpreting the observed asteroseismic data. 

With continuous photometric observations from Kepler, the frequency can be measured to the precision better than $10^{-6}$ day$^{-1}$. We can use the observables ($\delta$ and $\Delta_b+\Delta_c$) in the three-mode couplings to constrain the mode damping rate and thus potentially refine stellar parameters. Conversely, if we can first identify the modes and have accurate stellar parameters, we can predict the phase detuning of non-harmonic TEOs. Besides KIC 3230227, we expect more discoveries of mode couplings from tens of {\it Kepler} heartbeat binaries with TEOs. The ongoing Transiting Exoplanet Survey Satellite (TESS) observations, except for those in the continuous viewing zone, do not have the required time span to perform this kind of analysis as the phase detuning and frequency detuning require relatively high precision.
Note that the two observed phase detuning parameters ($\delta$, or $\Delta\phi$) are close to the values when $\tan(2\delta)$ reaches infinity ($\delta=0.25, 0.50$, or $\Delta\phi=0, -0.25$), this could an observational bias and it would be interesting to see if observed phase detuning are always close to these two values in more systems. Also note that the frequency detuning parameter $\Delta_b+\Delta_c$ of the two triplets are also very close to each other (Table 2). In fact, they are the same within $2\sigma$. This may be a requirement for the five modes to settle into a steady state.

We can perform tidal asteroseismology by using both the linear TEOs and the nonlinear TEOs:
the linearly driven TEOs are pure orbital harmonics, and which harmonics are excited primarily depends on the orbital parameters and stellar parameters. Pulsation phases can be used to identify these TEOs, but the observed pulsation amplitudes (flux variations) depend very sensitively on the detuning $Nf_{orb}-\omega$ and only have limited capability of constraining stellar parameters. Indeed, Fuller (2017) developed a probabilistic approach to model the linear TEO amplitudes.

It is possible to further test the nonlinear equilibrium solution as derived in Appendix D of W12. We have done a simplified calculation of the observed mode amplitude ratios. Ideally, we could direct test the theoretical daughter-mode amplitude relations (D12, D9) in W12 for the two triplets. We can also examine the three-mode parametric instability threshold and compare with observed flux variation of parent modes. This requires the mode coupling coefficient $\kappa$ to be calculated. This calculation is very demanding and has been performed for higher order g-modes in solar-type stars (W12; Weinberg \& Arras 2019) and white dwarfs (Wu \& Goldreich 2001; Luan \& Goldreich 2018) but rarely for early-type stars of spectral type F, A, or B. The exception is for KOI-54 and Burkart et al.\ (2012) found that the observed parent mode amplitude is lower than the three-mode coupling instability threshold. They argue that this is because the five-mode coupling lowers the threshold.

A caveat of this paper is that we have assumed the two observed mode triplets can be modeled as two independent $2^{nd}$-order three-mode couplings, and we only
integrate the AEs of three-mode coupling.  A more dedicated study should use the AEs of five coupled modes.  O'Leary \& Burkart (2014) have pioneered a study in this regard. But the rich behavior of multiple-mode coupling makes such a problem highly non-trivial. The full scanning of parameter space could potentially further constrain mode parameters. Such studies have not been done for early-type stars. We defer a more detailed study of the multiple mode coupling to a further paper.
 
\acknowledgments 
 
Z.G. is very grateful for the help from Phil Arras and Nevin Weinberg. We are in debt to the anonymous referee for his/her comments and suggestions. Some of this work was completed during my transit at the IAD airport. This work was partially supported by funding from the Center for Exoplanets and Habitable Worlds. The Center for Exoplanets and Habitable Worlds is supported by the Pennsylvania State University, the Eberly College of Science, and the Pennsylvania Space Grant Consortium.

 


\clearpage

\begin{deluxetable}{lccccccc} 
\tabletypesize{\footnotesize} 
\tablewidth{0pc} 
\tablenum{1} 
\tablecaption{Oscillation Frequencies \label{tab1}} 
\tablehead{ 
\colhead{}   & 
\colhead{Frequency (d$^{-1}$)}      &
\colhead{Amplitude (mag)}      &
\colhead{Phase (rad/$2\pi$)} & 
\colhead{S/N} &
\colhead{$N=f/f_{orb}$} &
%
}
\startdata 
\hline
 & Main frequencies reported in Guo et al. (2017)  & $ $ & $ $ & $$ & $$ \\
\hline
$f_{  4}$ & $ 1.9697649\pm  0.0000010$ & $0.000338\pm 0.000003$ & $0.1647\pm0.0035$ & $ 229.9$ & $13.881143(36)$ \\
$f_{  10}$ & $ 2.9799474\pm  0.0000012$ & $0.000194\pm 0.000002$ & $0.8681\pm0.0042$ & $ 191.8$ & $21.000006(54)$ \\
$f_{  5}$ & $ 2.1285534\pm  0.0000016$ & $0.000198\pm 0.000002$ & $0.8920\pm0.0055$ & $ 144.8$ & $15.000142(40)$ \\
$f_{  6}$ & $ 2.4123514\pm  0.0000016$ & $0.000177\pm 0.000002$ & $0.3867\pm0.0055$ & $ 144.3$ & $17.000096(45)$ \\
$f_{  8}$ & $ 2.6961494\pm  0.0000017$ & $0.000154\pm 0.000002$ & $0.3596\pm0.0057$ & $ 139.0$ & $19.000051(50)$ \\
$f_{  2}$ & $ 1.7198829\pm  0.0000020$ & $0.000192\pm 0.000003$ & $0.3419\pm0.0069$ & $ 115.4$ & $12.120198(34)$ \\
$f_{  7}$ & $ 2.5542504\pm  0.0000022$ & $0.000124\pm 0.000002$ & $0.8654\pm0.0075$ & $ 106.9$ & $18.000074(48)$ \\
$f_{  1}$ & $ 1.4021347\pm  0.0000026$ & $0.000179\pm 0.000003$ & $0.9722\pm0.0089$ & $  89.2$ & $ 9.880992(31)$ \\
$f_{  9}$ & $ 2.8380484\pm  0.0000034$ & $0.000073\pm 0.000002$ & $0.3470\pm0.0116$ & $  69.0$ & $20.000028(56)$ \\
$f_{ 3}$ & $ 1.8448239\pm  0.0000043$ & $0.000085\pm 0.000003$ & $0.3183\pm0.0147$ & $  54.4$ & $13.000670(45)$ \\
\hline
 & Additional Frequencies  & $ $ & $ $ & $$ & $ $ \\
\hline
$f_{ 11}$ & $ 3.1218608\pm  0.0000053$ & $0.000043\pm 0.000002$ & $0.3299\pm0.0180$ & $  44.4$ & $22.000085(67)$ \\
$f_{ 12}$ & $ 1.7027879\pm  0.0000057$ & $0.000069\pm 0.000003$ & $0.4935\pm0.0195$ & $  40.9$ & $11.999727(51)$ \\
$f_{ 13}$ & $ 3.4056245\pm  0.0000064$ & $0.000033\pm 0.000002$ & $0.3161\pm0.0218$ & $  36.7$ & $23.999798(76)$ \\
$f_{ 14}$ & $ 3.2637065\pm  0.0000071$ & $0.000031\pm 0.000002$ & $0.3312\pm0.0243$ & $  32.9$ & $22.999686(77)$ \\
$f_{ 15}$ & $ 3.6894194\pm  0.0000083$ & $0.000024\pm 0.000001$ & $0.3187\pm0.0284$ & $  28.1$ & $25.999731(89)$ \\
$f_{ 16}$ & $ 1.8441372\pm  0.0000086$ & $0.000042\pm 0.000003$ & $0.5133\pm0.0293$ & $  27.2$ & $12.995831(69)$ \\
$f_{ 17}$ & $ 4.3989829\pm  0.0000107$ & $0.000016\pm 0.000001$ & $0.0207\pm0.0365$ & $  21.8$ & $31.000000(109)$ \\
$f_{ 18}$ & $ 3.9732631\pm  0.0000110$ & $0.000017\pm 0.000001$ & $0.3246\pm0.0376$ & $  21.2$ & $28.000007(106)$ \\
$f_{ 19}$ & $ 2.2703785\pm  0.0000111$ & $0.000027\pm 0.000002$ & $0.4319\pm0.0380$ & $  21.0$ & $15.999599(88)$ \\
$f_{ 20}$ & $ 3.8313321\pm  0.0000114$ & $0.000017\pm 0.000001$ & $0.3228\pm0.0391$ & $  20.4$ & $26.999804(106)$ \\
$f_{ 21}$ & $ 1.4189884\pm  0.0000128$ & $0.000036\pm 0.000003$ & $0.5079\pm0.0438$ & $  18.2$ & $ 9.999762(94)$ \\
$f_{ 22}$ & $ 0.7094950\pm  0.0000133$ & $0.000065\pm 0.000006$ & $0.6236\pm0.0455$ & $  17.5$ & $ 4.999886(95)$ \\
$f_{ 23}$ & $ 1.9866491\pm  0.0000136$ & $0.000025\pm 0.000002$ & $0.2494\pm0.0465$ & $  17.1$ & $14.000127(103)$ \\
$f_{ 24}$ & $ 5.6760898\pm  0.0000138$ & $0.000010\pm 0.000001$ & $0.8318\pm0.0470$ & $  17.0$ & $40.000008(141)$ \\
$f_{ 25}$ & $ 4.2570230\pm  0.0000197$ & $0.000009\pm 0.000001$ & $0.4460\pm0.0672$ & $  11.9$ & $29.999693(159)$ \\
$f_{ 26}$ & $ 2.2892101\pm  0.0000220$ & $0.000014\pm 0.000002$ & $0.0677\pm0.0751$ & $  10.6$ & $16.13231(160)$ \\
$f_{ 27}$ & $ 1.5606910\pm  0.0000234$ & $0.000018\pm 0.000003$ & $0.8303\pm0.0800$ & $  10.0$ & $10.99836(168)$ \\
\hline
$f_{      orb}$    & $  0.1419022  \pm 0.0000004$ & $ -$ & $ -$ & $-$ & $-$ \\
\enddata 
\end{deluxetable}

\begin{deluxetable}{lccccccc} 
\tabletypesize{\footnotesize} 
\tablewidth{0pc} 
\tablenum{2} 
\tablecaption{Resonant Three-mode Coupling $f_{b}+f_{c} \approx f_{a}$ Observables \label{tab2}} 
\tablehead{ 
\colhead{}   & 
\colhead{Frequency $f_i$ (day$^{-1}$)}      &
\colhead{Amplitude $a_i$ (mag)}      &
\colhead{Phase $\phi_i$ (rad/$2\pi$)} & 
\colhead{$N=f/f_{orb}$} &
}
\startdata 
$f_{orb}$ (Orbital Frequency)& $0.14190222(36)$ & $ $ & $$ & $ $ \\
\hline
Triplet 1 ($f_{2}+f_{1} \approx f_{11}$)& $ $ & $ $ & $$ & $$ \\
\hline
$f_{b}$ (daughter)$ : f_{  2}$ & $ 1.7198829(20)$ & $0.000192(3)$ & $0.3419(69)$  & $12.120198(34)$ \\
$f_{c}$ (daughter)$ : f_{  1}$ & $ 1.4021347(26)$ & $0.000179(3)$ & $0.9722(89)$  & $ 9.880992(31)$ \\
$f_{a}$ (parent)$ :f_{ 11}$ & $ 3.1218608(53)$ & $0.000043(2)$ & $0.3299(180)$  & $22.000085(67)$ \\
$\Delta \phi $ $= \phi_{11}-\phi_{2}-\phi_{1}$  & $ $ & $ $ & $0.0158\pm 0.0210$ & $ $ \\
$\delta=\pi/2-\Delta\phi$& $ $ & $ $ & $0.2342\pm 0.0210$ & $ $ \\
$\Delta_b+\Delta_c=22f_{orb}-f_{2}-f_{1}$ & $-0.0001687(86) $ & $ $ & $$ & $ $ \\
$\gamma_b+\gamma_c$ (model dependent) & $\approx 10^{-7}-10^{-4}$ & $ $ & $$ & $ $ \\
\hline
Triplet 2  ($f_{  4}+f_{2} \approx f_{15}$) & $ $ & $ $ & $$ & $ $ \\
\hline
$f_{b}$ (daughter): $f_{  4}$  & $ 1.9697649(10)$ & $0.000338(3)$ & $0.1647(35)$  & $13.881143(36)$ \\
$f_{c}$ (daughter): $f_{  2}$  & $ 1.7198829(20)$ & $0.000192(3)$ & $0.3419(69)$  & $12.120198(34)$ \\
$f_{a}$ (parent): $f_{ 15}$ & $ 3.6894194(83)$ & $0.000024(1)$ & $0.3187(284)$  & $25.999731(89)$ \\
$\Delta\phi=  \phi_{15}-\phi_{2}-\phi_{4}$  & $ $ & $ $ & $ -0.1879\pm 0.0294$ & $ $ \\
$\delta=\pi/2-\Delta\phi$& $ $ & $ $ & $0.4379\pm 0.0294$ & $ $ \\
$\Delta_b+\Delta_c=26f_{orb}-f_{2}-f_{4}$ & $-0.0001900(97)$ & $ $ & $$ & $ $ \\
$\gamma_b+\gamma_c$ (model dependent) & $\approx 10^{-7}-10^{-4} $ & $ $ & $$ & $ $ \\
\enddata 
\end{deluxetable}



\begin{figure} 
\begin{center} 
{\includegraphics[angle=0,height=17cm,width=13cm]{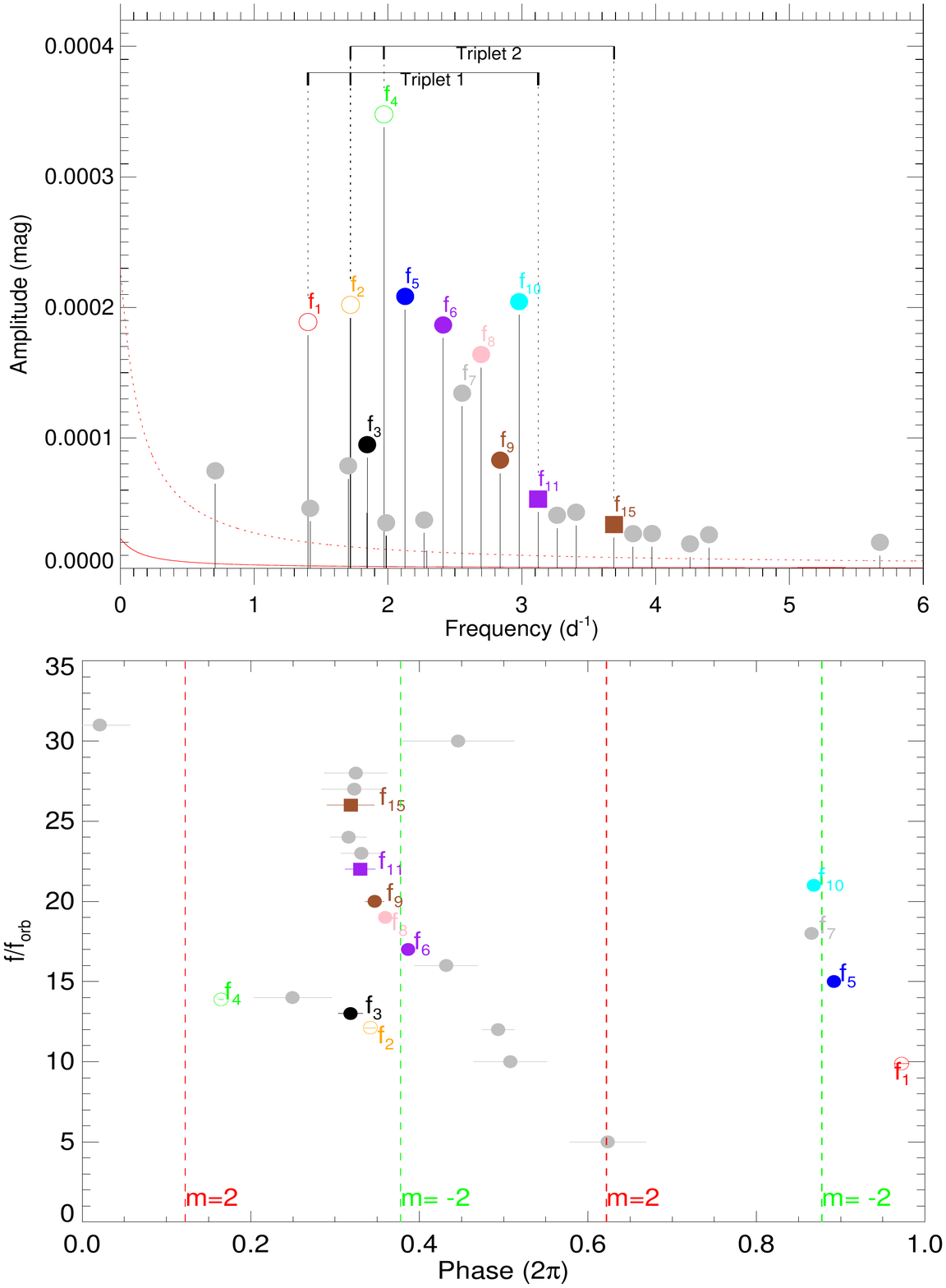}} 
\end{center} 
\caption{\textbf{Upper:} Fourier amplitude spectrum of KIC 3230227 after subtracting the binary star light curve. The 10 dominant TEOs reported in Guo, Gies \& Fuller (2017) are labeled as $f_1-f_{10}$. The newly extracted frequencies with $S/N > 10$ are marked with gray symbols. The solid and dotted red lines represent  the $1\sigma$ and $10\sigma$ noise model. The two triplets are indicated by black dotted lines and the two `forks' on top of the frequency peaks. \textbf{Lower:} Pulsation phases of TEOs. The newly extracted frequencies are shown as gray dots. Theoretical phases of $m=2$ and $m=-2$ quadruple modes, if these are tidally forced oscillations, are indicated by the red and green dashed lines, respectively. }
\end{figure}

\begin{figure} 
\begin{center} 
{\includegraphics[angle=0,height=12cm]{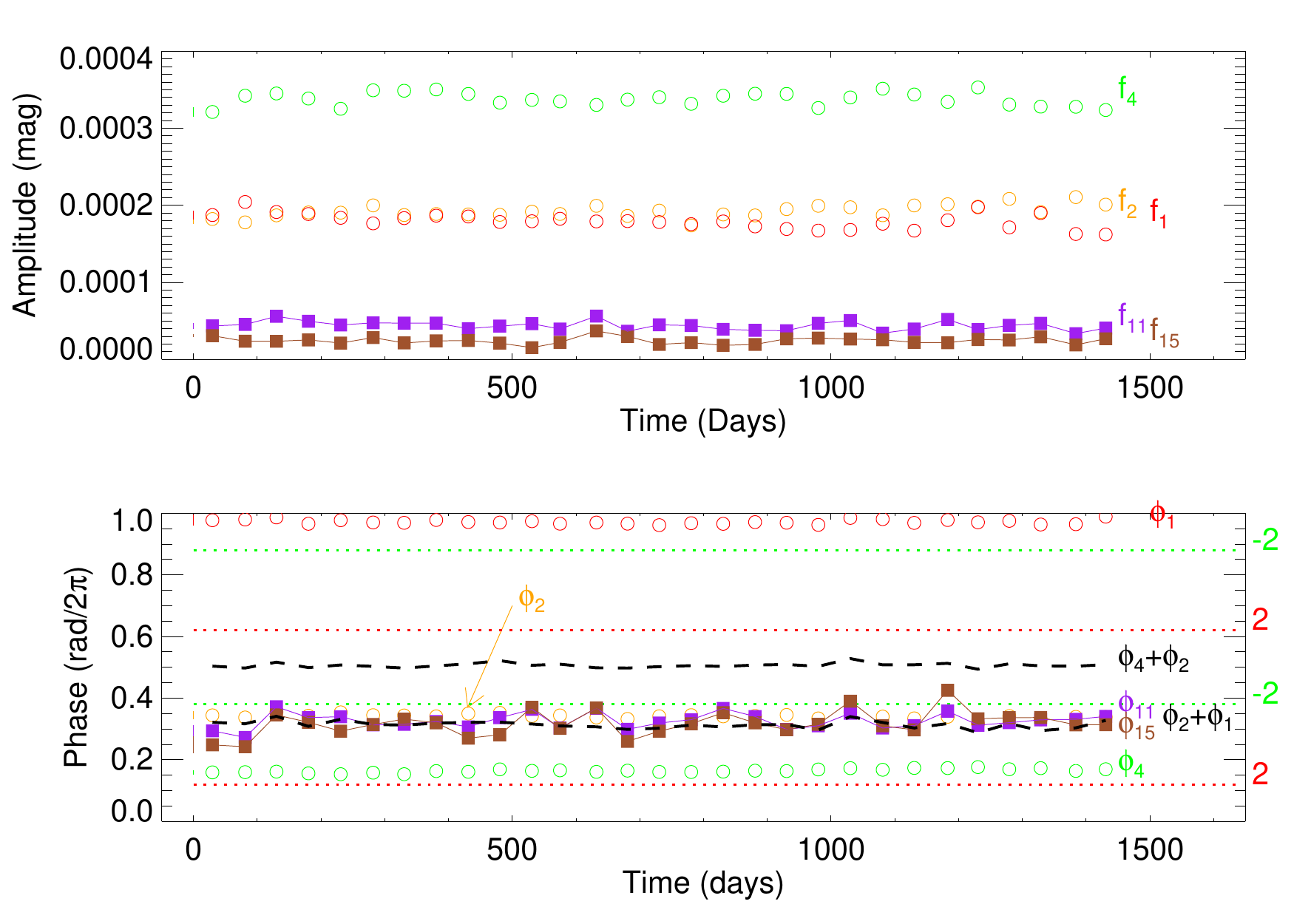}} 
\end{center} 
\caption{\textbf{Upper}: Amplitude variations of the daughter modes ($f_4,f_2,f_1$) and parent modes ($f_{15}, f_{11}$). \textbf{Lower}: The corresponding phase variations: ($\phi_4, \phi_2, \phi_1$) and ($f_{15},f_{11}$). The vertical line segments at Time $=0$ indicate the typical $2\sigma$ error bars. The theoretical phases for $(l=2,m=2)$ and $(l=2,m=-2)$ are marked by the horizontal dotted lines in red and green, respectively. The sum of the two daughter modes phases in the two triplets ($\phi_4+\phi_2$ and $\phi_1+\phi_2$) are also plotted, and they can be compared to their parents' phases $\phi_{15}$ and $\phi_{11}$. It can be seen that the amplitudes and phases are almost stable over the time span of {\it Kepler} observations ($4$ yrs), and the phases satisfy the following relations:  $\phi_{15}-(\phi_4+\phi_2) \approx -0.2$  and $\phi_{11}-(\phi_2+\phi_1) \approx 0 $.}
\end{figure}

\begin{figure} 
\begin{center} 
{\includegraphics[angle=0,height=10cm]{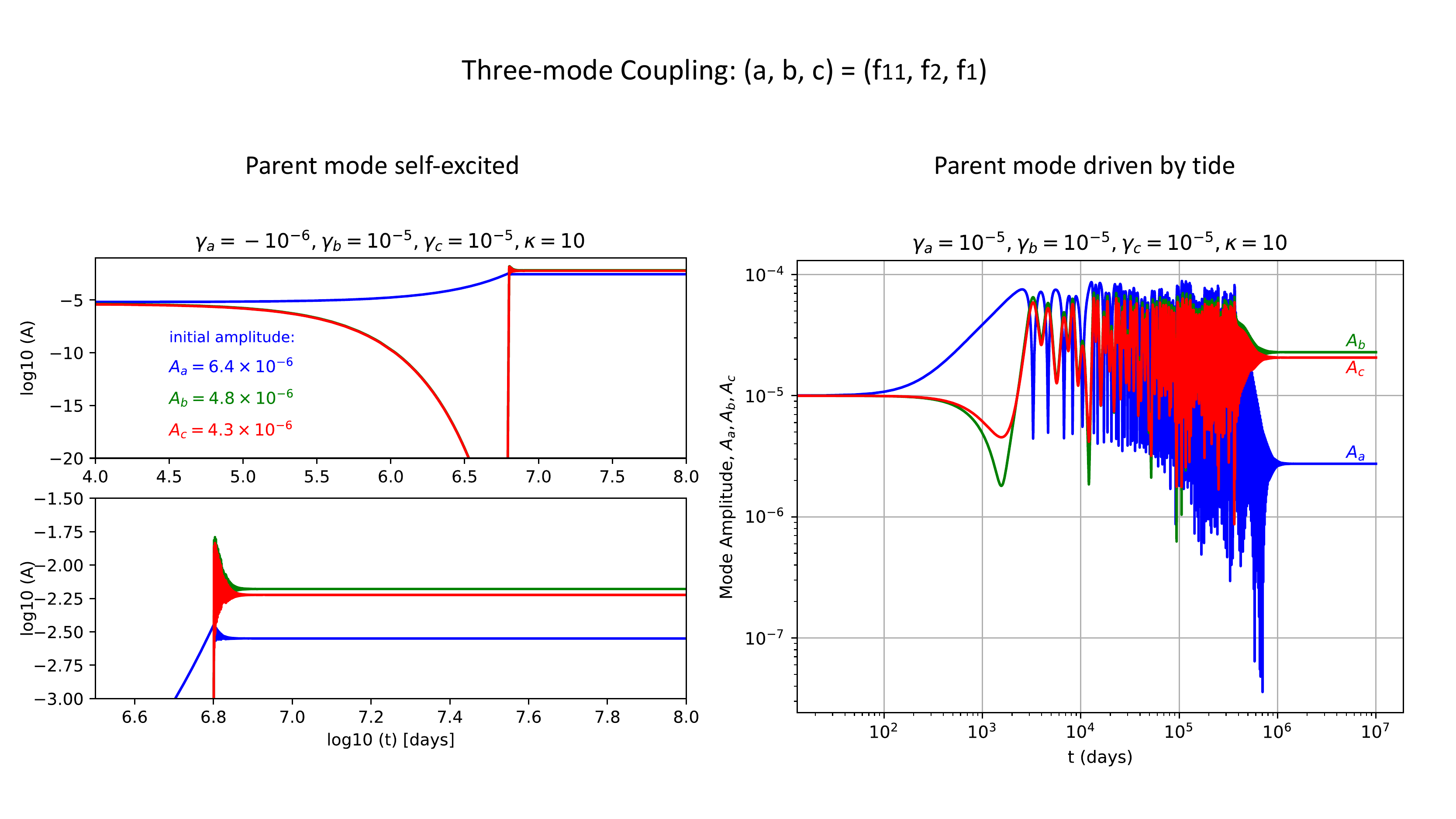}} 
\end{center} 
\caption{Mode amplitude evolution of a three-mode system representing the observed g-mode triplet in KIC 3230227: $(f_a, f_b, f_c)=(f_{11}, f_2, f_1).$ \textbf{Left:} Parent mode is self-excited, with $\gamma_a <0$; \textbf{Right:} Parent mode is stable ($\gamma_a >0$), but driven by a tidal term $U_a e^{-i\omega t}$, with $\omega$ being an orbital harmonic ($=22 f_{orb}$). In both cases, the three-mode system settles into an equilibrium state.}
\end{figure} 

\begin{figure} 
\begin{center} 
{\includegraphics[angle=0,height=13cm]{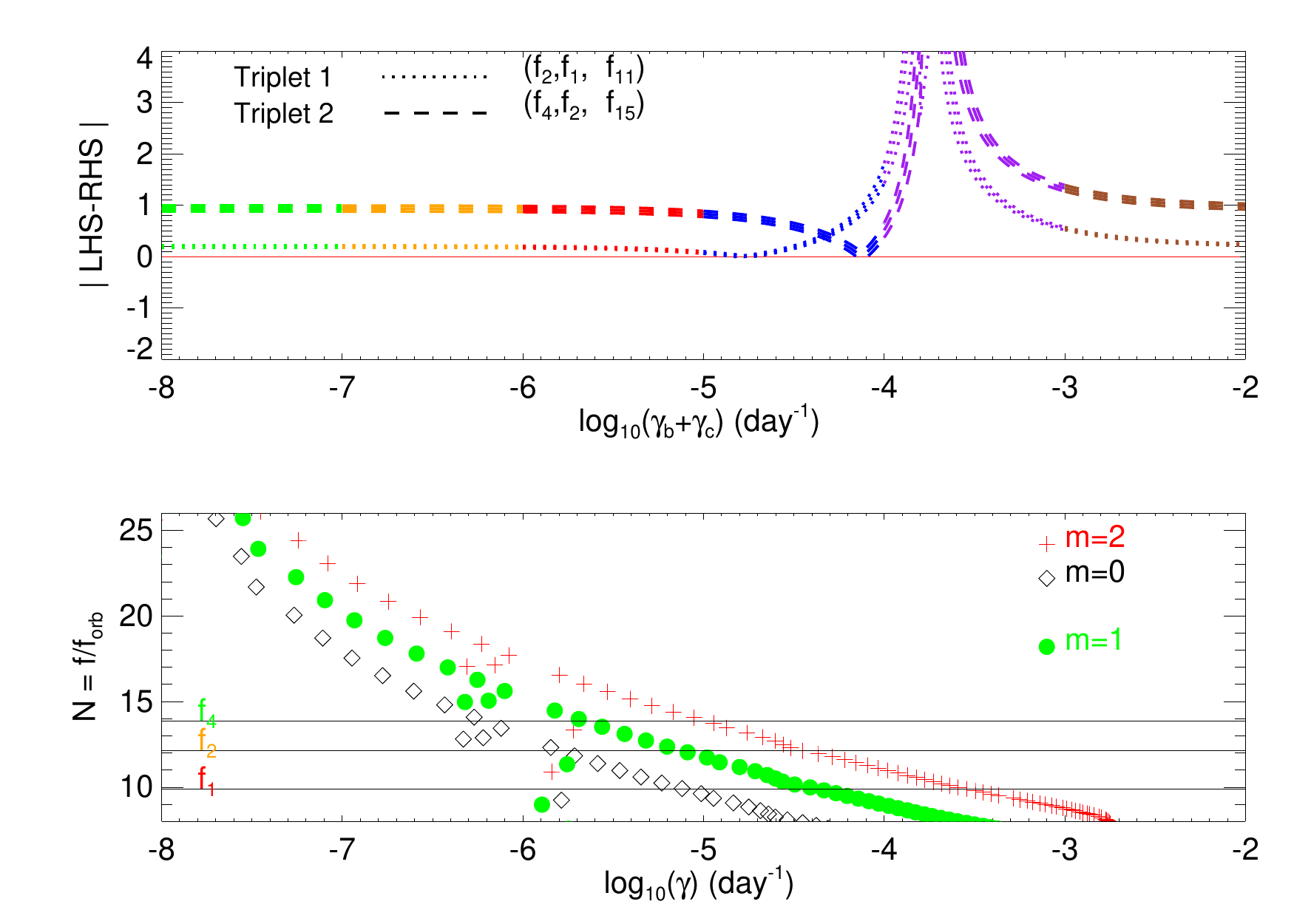}} 
\end{center} 
\caption{\textbf{Upper:} The difference between LHS and RHS of \textbf{eq.3} is shown as a function of the sum of the mode damping rates of two daughter modes. \textbf{Lower:} The linear damping rate of $l=2$ g-modes calculated with GYRE from a representative stellar model with $M=1.84M_{\odot}$, $R=2.01R_{\odot}$. In the non-adiabatic calculation, rotation is treated in the traditional approximation. The observed frequencies of three daughter modes ($f_1, f_2, f_4$) are indicated by the horizontal lines.}
\end{figure}

\end{document}